\documentclass[aps]{revtex4}
\input{epsf}
\begin{document}
\title{Spiral phase and phase separation of the double exchange model
in the large-S limit}
\author{Lan Yin}
\email{yinlan@pku.edu.cn} \altaffiliation[Former
address:]{Department of Physics, University of Washington,
Seattle, WA 98195-1560} \affiliation{School of Physics, Peking
University, Beijing 100871, P. R. China}

\date{\today}

\begin{abstract}
The phase diagram of the double exchange model is studied in the large-S limit at
zero temperature in two and three dimensions.  We find that the spiral state has
lower energy than the canted antiferromagnetic state in the region between the
antiferromagnetic phase and the ferromagnetic phase.  At small doping, the spiral
phase is unstable against phase separation due to its negative compressibility.
When the Hund coupling is small, the system separates into spiral regions and
antiferromagnetic regions.  When the Hund coupling is large,  the spiral phase
disappears completely and the system separates into ferromagnetic regions and
antiferromagnetic regions.
\end{abstract}
\maketitle

The colossal magnetoresistance effect(CMR) was discovered in
hole-doped manganese oxides such as $\rm La_{1-x}Sr_xMnO_3$ and
$\rm La_{1-x}Ca_xMnO_3$ \cite{Intro}.  Various experiments have
revealed that these materials have very rich phase
diagrams~\cite{PD}.  In these materials, the three $t_{2g}$
electrons form a localized $S=3/2$ Kondo spin at each manganese
site and $e_g$ electrons form a conduction band.  The degeneracy
of the two $e_g$ orbitals is lifted by Jahn-Teller effect and only
one of the orbitals is close to the fermi energy. The $e_g$ spins
interact with $t_{2g}$ spins through Hund coupling.  The single
orbital double exchange (DE) model\cite{DEM} is the simplest
description of this system.  There are other important factors in
this system, such as Jahn-Teller phonons, exchange interaction
between the Kondo spins, and Columb interaction.  It is important
to study the DE model and identify its role in the CMR systems.

The ferromagnetic phase of the DE model have been studied in the past
\cite{DEM,DEFM}.  De Gennes studied the DE model with the exchange interaction
and found that the system goes into a canted antiferromagnetic (CAF) phase near
half filling~\cite{CAF}.  However numerical studies~\cite{NUM} discovered phase
separation in the simple DE model near half filling in one, two and infinite
dimensions.  There is some evidence that it occurs in three dimensions as well.
In addition, an incommensurate phase was found when Hund coupling is relatively
small.  Phase separation was also found in DE models with exchange
interaction and with Jahn-Teller phonons~\cite{MNUM}.

Some theoretical studies~\cite{FTPS,TJM,ES,TP,CMD} have been focused on phase
separation in the DE model.  Phase separation was also found in the limit of
infinite Hund coupling near Curie temperature, but absent in zero temperature
\cite{FTPS}.  The DE model was mapped onto a $t-J$ model in Ref.~\cite{TJM} where
phase separation was found not only near half filling~\cite{PSTJ}, but also near
zero filling.  The stability of canted
antiferromagnetic state was examined in Ref.~\cite{ES} and phase separation was
found in certain parameter regions, but it is unclear that without exchange
interaction whether the phase separation exists or not.  In Ref.~\cite{TP} and
Ref.~\cite{CMD}, phase separation was found in the DE model within the dynamic
mean field approximation.

In this paper, we study the zero-temperature phase diagram of the
simple DE model in the large-S limit in cubic and square lattices.
At zero doping, the system is in an antiferromagnetic phase.  At a
high level of doping, the system goes into a ferromagnetic phase.
We find that in the region between the antiferromagnetic phase and
the ferromagnetic phase, the spiral state has the lower energy
than the CAF state. However, the spiral state is always unstable
near half filling and subject to phase separation.  When the ratio
of Hund coupling to hopping energy is below certain critical
value, the separation is between the antiferromagnetic phase and
the spiral phase.  Above this ratio, the spiral phase is always
unstable and the system always separates into antiferromagnetic
phase and ferromagnetic phase.  There is one qualitative
difference between 2D and 3D systems.  In 3D, in the limit of zero
Hund coupling, the transition between the ferromagnetic phase and
the spiral phase occurs at hole density around 0.55; in the same
limit, the critical density approaches 1 in 2D.

The Hamiltonian of the DE model is given by
\begin{equation}
{\cal H}=-t \sum_{\langle ij \rangle,\sigma} c_{i \sigma}^\dagger c_{j \sigma}
-J \sum_i {\bf S}_i \cdot {\bf s}_i,
\end{equation}
where ${\bf s}_i$ is the electron spin with total spin $1/2$ and ${\bf S}_i$ is
the Kondo spin with total spin $S$.  In this paper, we consider the Kondo spins
as classical spins which is exact in the large-$S$ limit.  The quantum correction
to this approximation is smaller by a factor of $1/S$ and vanishes in the large-$S$
limit.  The classical spin approximation is
also equivalent to trial wave-functions with static spin configuration.  Thus it
can provide an upper bound for ground state energy.

We consider the possible phases of the system.  For convenience,
we use the electron operators $f_i$ and $h_i$ which have spins
parallel and opposite to the Kondo spin,
\begin{eqnarray*}
c_{i\uparrow} &=& \cos(\theta_i/2) e^{i\alpha_i} f_i+
\sin(\theta_i/2) e^{-i(\alpha_i+\phi_i)} h_i,\\
c_{i\downarrow} &=& -\cos(\theta_i/2) e^{-i\alpha_i} h_i+
\sin(\theta_i/2) e^{i(\alpha_i+\phi_i)} f_i,
\end{eqnarray*}
where $\alpha_i$ is an arbitrary phase factor, and $\theta_i$ and
$\phi_i$ are the azithmuth and polar angles of ${\bf S}_i$.  For
the Kondo spins, it is convenient to use the Schwinger-boson
representation~\cite{SB}, $b_{i\uparrow}=\sqrt{2S}
\cos(\theta_i/2) e^{i\alpha_i}$ and
$b_{i\downarrow}=\sqrt{2S}\sin(\theta_i/2) e^{i(\phi_i+\alpha_i)}$
with the constraint $\sum_{\sigma} b_{i\sigma}^* b_{i\sigma}=2S$.
In the classical spin approximation, the Schwinger-boson operators
are just complex numbers.  In terms of the new fermion operators
and the Scwinger-boson operators, the Hamiltonian is given by
\begin{eqnarray}
{\cal H} &=& -{t \over 2S}\sum_{\langle ij\rangle}
\left[(f_i^\dagger f_j+h_j^\dagger h_i) \sum_\sigma b_{i\sigma}^* b_{j\sigma}
+\left(f_i^\dagger h_j (b_{i\uparrow}^* b_{j\downarrow}^* -b_{j\uparrow}^*
b_{i\downarrow}^*)+ h.c. \right)\right] \nonumber \\
& &-{JS \over 2}\sum_i (f_i^\dagger f_i- h_i^\dagger h_i).
\end{eqnarray}

In the ferromagnetic phase, all the Kondo spins are aligned in the
same direction, and the electron hopping amplitude is diagonal and
at maximum, $\sum_\sigma b_{i\sigma}^* b_{j\sigma}=2S$ and
$b_{i\uparrow} b_{j\downarrow} -b_{j\uparrow} b_{i\downarrow}=0$.
The $h$- and $f$-fermions are free particles with the dispersion
given by $E_{\bf k}=\epsilon_{\bf k}\pm{1 \over 2}JS$, where
$\epsilon_{\bf k}= -2t\sum_{\alpha=1}^d \cos k_\alpha a$, a is the
lattice spacing and $d$ is the dimensionality.  When the electron
bandwidth $4dt$ is bigger than the Hund energy splitting $JS$, for
less than half filling, only the lower band is occupied and this
fully-magnetized state is an exact eigenstate the Hamiltonian.
When $JS<4dt$ and sufficiently close to half filling, both bands
become occupied and the ferromagnetic state is no longer an exact
eigenstate but an approximation. It is straight forward to go
beyond classical spin approximation and show that the spin
fluctuations are essentially described spin waves with quadratic
dispersion~\cite{DEM,DEFM}, much like in a typical Heisenberg
ferromagnet.  At zero temperature, the spin waves are frozen out
and do not provide any significant change to the classical spin
configuration.

We have also considered the possibility of a totally disordered state.
However, in the classical spin approximation it is insufficient to consider
such a state.  We use a mean-field Schwinger-boson formalism~\cite{SB}
instead.  But we found that the saddle point of the disordered state
does not exist at zero temperature for any positive integer or half-integer
Kondo spin in dimensions higher than or equal to two.

Exactly at half filling, the system is in an antiferromagnetic
phase. For simplicity, we choose the Kondo spins aligning in the
$x$-direction, $b_{i\uparrow}=\sqrt{S}e^{-i {\bf Q} \cdot {\bf
R}_i}$ and $b_{i\downarrow}=\sqrt{S} e^{i{\bf Q} \cdot {\bf
R}_i}$, where ${\bf Q}= ({\pi \over 2a},{\pi \over 2a},{\pi \over
2a})$.  The hopping matrix becomes completely off-diagonal,
$\sum_\sigma b_{i\sigma}^* b_{j\sigma}=0$ and $b_{i\uparrow}
b_{j\downarrow} -b_{j\uparrow} b_{i\downarrow}= -2iS\sin\left({\bf
Q} \cdot ({\bf R}_i-{\bf R}_j)\right)$.  The fermions form two
bands with dispersion given by $E_{\bf k}=\pm\sqrt{{J^2S^2 \over
4}+ {\epsilon^\prime}^2_{\bf k}}$, where ${\epsilon^\prime_{\bf
k}= -2t\sum_{\alpha}\sin k_\alpha a}$.  The free-fermion operators
are linear combinations of the $f$- and $h$-operators. At half
filling, only the lower band is occupied.

In comparison, when $JS>4dt$, the lower band of the ferromagnetic state are
completely filled and the total hopping energy is zero at half filling.
It is clear that the antiferromagnetic phase has lower energy in this case.
When $JS<4dt$, the total energy
difference between these two states is given by
\begin{equation} \label{Dif}
\Delta E=\int_{-{1 \over 2}JS}^{{1\over 2}JS}(\sqrt{({1\over2}JS)^2+\epsilon^2}
-{1\over 2}JS) N(\epsilon) d\epsilon+\int_{{1\over 2}JS}^{2dt}
2(\sqrt{({1\over2}JS)^2+\epsilon^2}-\epsilon) N(\epsilon) d\epsilon,
\end{equation}
where $N(\epsilon) \equiv \sum_{\bf k} \delta(\epsilon_{\bf k}-\epsilon)=
\sum_{\bf k} \delta(\epsilon_{\bf k}^\prime-\epsilon)$.  The ferromagnetic
state has higher energy in this case as well.

It is not difficult to go beyond the classical spin approximation to show that
there are linearly dispersed spin waves in the antiferromagnetic state.  Here
the quantum fluctuations are stronger because the classical antiferromagnetic
spin state is not an eigenstate of the Hamiltonian.  The actual staggered
magnetization is smaller due to fluctuations.  But it does not change the
fact that the antiferromagnetic state has lower energy since the true
antiferromagnetic state has even lower energy than the classical state.

The properties of ferromagnetic state are totally different from those of the
antiferromagnetic state.  It is not a total surprise that other states have
lower energy in the region between the two phases, as found in Ref.~\cite{NUM}.
Here we consider two homogeneous states which are natural candidates, the spiral
state and the CAF state.

The spiral state was considered as a possible description of doped
high Tc systems\cite{SPR}, although few supportive experimental
evidence have been found so far.  The spiral state is a compromise
between the hopping of the holes and the effective superexchange
interaction of the electrons. The ferromagnetic and
antiferromagnetic phases are just two extreme cases of the spiral
phase. The possibility of a spiral phase was considered in the DE
model with exchange interaction in the large Hund coupling
limit~\cite{DESPR}.  It was found that there is a transition from
the ferromagnetic phase to the spiral phase near zero doping as a
function the of exchange coupling.  However, it was not addressed
whether the spiral phase exists or not in the simple DE model with
arbitrary Hund coupling and without any exchange interaction.

In a spiral state, the Kondo spin is aligned in a certain plane
and the angle of the spin is a linear function of its site
position vector.  For simplicity, we consider the spins in the
$x$-$y$ plane and choose the classical Schwinger-boson operators
to be $b_{i\uparrow}=\sqrt{S}e^{-i {\bf Q} \cdot {\bf R}_i}$ and
$b_{i\downarrow}=\sqrt{S} e^{i{\bf Q} \cdot {\bf R}_i}$. The
Hamiltonian is now given by
\begin{eqnarray}
{\cal H} &=& -\sum_{\langle ij\rangle}\left[\cos\left({\bf Q} \cdot
({\bf R}_i-{\bf R}_j) \right)(f_i^\dagger f_j+h_j^\dagger h_i)
+ i \sin\left({\bf Q} \cdot ({\bf R}_i-{\bf R}_j)\right)(f_i^\dagger h_j-
h_j^\dagger f_i)\right] \nonumber \\
& & -{JS \over 2}\sum_i(f_i^\dagger f_i-h_i^\dagger h_i).
\end{eqnarray}
For symmetry reasons, we consider the ${\bf Q}$-wavevector in $(1,1,1)$ direction
which is equivalent to all the other $(\pm 1,\pm 1,\pm 1)$ directions.
After diagonalizing this Hamiltonian, we obtain two free fermion bands
with dispersion given by
\begin{equation}
E_{\bf k}=\cos(Q_x a)\epsilon_{\bf k} \pm \sqrt{{J^2S^2 \over 4}+
\sin(Q_x a)^2 {\epsilon^\prime}^2_{\bf k}}.
\end{equation}
The ferromagnetic phase is a special case with ${\bf Q}=0$ and
the antiferromagnetic phase is another special case with $Q_x=\pi/2$.
The wavevector ${\bf Q}$ of the preferred spiral
state can be obtained by minimizing the total energy.

In a CAF state, each sublattice has different magnetization directions.
The uniform magnetization coexists with the staggered magnetization.
The ferromagnetic and antiferromagnetic states are also two special cases
of the CAF states.  To study the CAF state, we take the $z$-direction to be
the direction of the uniform component and $x$-direction to be the direction
of the staggered component, with the $b$-values given by $b_{i\uparrow}=b_1$
and $b_{i\downarrow}=(-1)^i b_2$.  The Hamiltonian can be easily diagnalized.
The fermions form two bands with dispersion given by
\begin{equation}
E_{\bf k}=\pm\sqrt{({JS\over 2})^2+\epsilon_{\bf k}^2+JS_z\epsilon_{\bf k}},
\end{equation}
where the uniform magnetization is given by $S_z=(|b_1|^2-|b_2|^2)/2$.

\begin{figure}
\epsfxsize=12cm
\centerline{ \epsffile{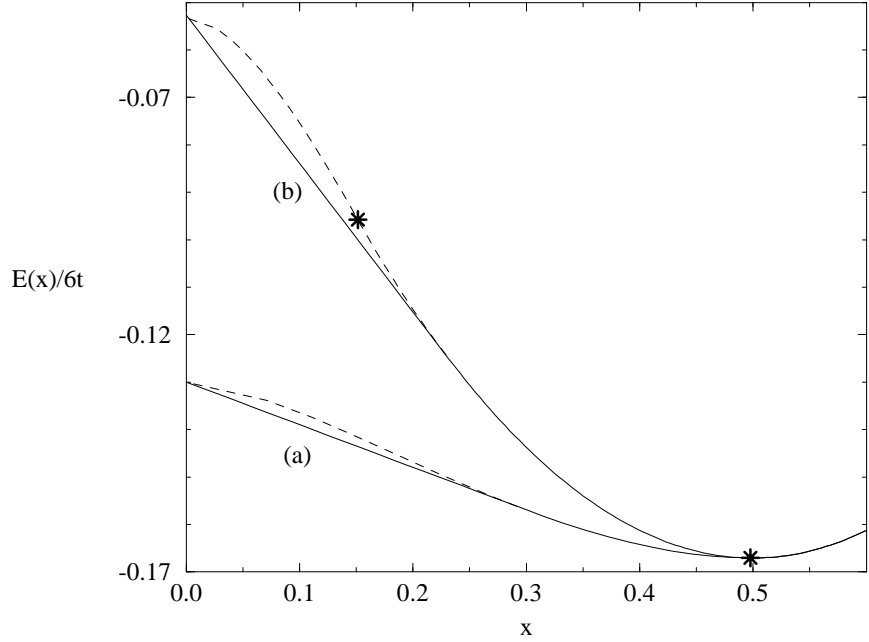} }
\vspace{0.3cm}
\caption{\label{fig1}
The energies of the 3D DE model in the large-S limit as a function of hole density
$x$ for (a) $JS=6t$ and (b) $JS=18t$.  $E(x)$ is the total energy per site subtracted
by chemical potential $-{1 \over 2}(1-x)JS$.  The solid lines are the lowest energies
of the system.  In (a), the dotted line is the energy of the spiral state
in the phase separation region.  The energy of phase separation is given by
the straight line below it.  The star marks the transition point from the
ferromagnetic phase to the spiral phase.  In (b), the dotted line is the
energy of the spiral state in the phase separation region.  The star marks the
hypothetical transition point between the spiral phase and the ferromagnetic phase.}
\end{figure}

We compared the energies of different states.  The CAF states
always have higher energy than the spiral state.  In
Fig.~\ref{fig1}, the energy of the 3D DE model is plotted as a
function of hole density. As shown in this figure, the
ferromagnetic phase always have the lowest energy when the hole
density is above certain value; below this density, the spiral
phase has lower energy.  At half filling, the antiferromagnetic
phase has the lowest energy.  However, near half filling, as shown
in Fig.~\ref{fig1}, the spiral phase is always unstable due to its
negative compressibility and it is subject to phase separation.
One of the phases in phase separation is the antiferromagnetic
phase.  Depending on the value of $JS/4dt$, the other phase can be
either ferromagnetic phase or the spiral phase.

To determine the phase boundary where phase separation occurs we use
in the following method.  Let x be the overall hole density of the system
and $y$ be the ratio of the volume of the ferromagnetic regions or
the spiral regions to the total system volume, the total energy of the
system per site is given by
\begin{equation}
E(x,y)=(1-y)E_A+yE_B({x \over y}),
\end{equation}
where $E_A$ is the energy of the antiferromagnetic phase per site and
$E_B(x)$ is the energy of the ferromagnetic or the spiral phase per site
at hole density $x$.  Since the energy $E(x,y)$ is at minimum as a
function of $y$, the stability condition $\partial_y E(x,y)=0$ produces
the equation
\begin{equation}\label{EQY}
E_A=E_B({x \over y}) -{x \over y}E_B'({x \over y}).
\end{equation}
The solution of eq.~(\ref{EQY}) is given by
\begin{equation}\label{SCL}
{x \over y}=x_c,
\end{equation}
where the constant $x_c$ is the critical hole density of the
system since the phase boundary is determined by $y=1$.  The hole
density of the spiral regions or ferromagnetic regions $x/y$ is
always equal to the critical hole density $x_c$.

We find that the system separates into the spiral phase and the
antiferromagnetic phase when ${JS \over 4dt}$ is smaller than certain
value.  When ${JS \over 4dt}$ is bigger than this critical value, the spiral phase
disappears completely and the phase separation is between the ferromagnetic
phase and the antiferromagnteic phase.  In 3D, the critical ratio is about
$2/3$; it is about $0.84$ in 2D.  In both cases, the critical hole density of
phase separation boundary vanishes in the limit of $J \rightarrow \infty$ and
also in the limit of $J \rightarrow 0$.

\begin{figure}
\epsfxsize=12cm
\centerline{ \epsffile{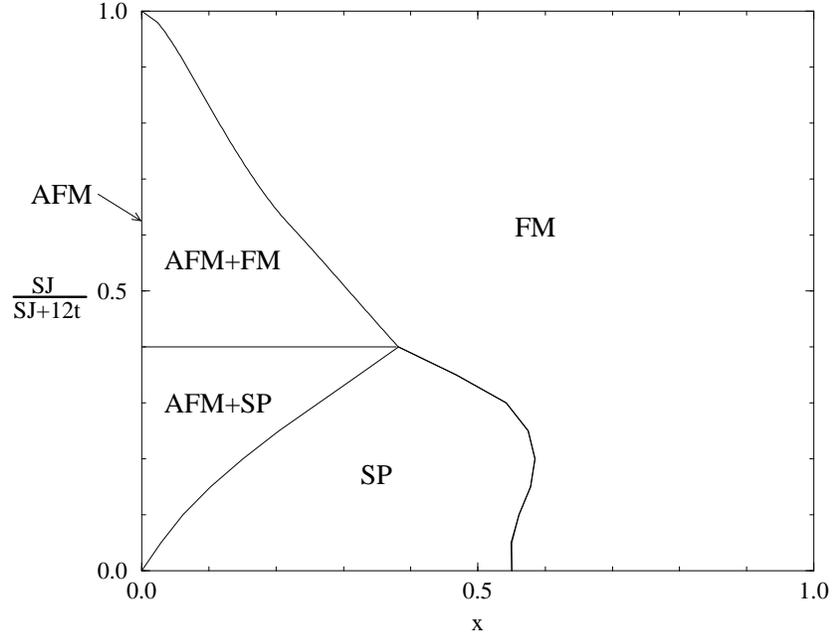} }
\vspace{0.3cm}
\caption{\label{fig2}
The zero temperature phase diagram of double exchange model
in cubic lattice in the large-S limit.  The ferromagnetic phase (FM) exists
at high hole density and the antiferromagnetic phase (AFM) exists at half
filling.  When ${JS \over 12 t}<{2 \over 3}$, the spiral phase (SP) appears
at low hole density.  Phase separation between SP and
AFM occurs near zero doping.  When ${JS \over 12 t}>{2 \over 3}$,
phase separation between FM and AFM appears at low doping.}
\end{figure}

\begin{figure}
\epsfxsize=12cm \centerline{ \epsffile{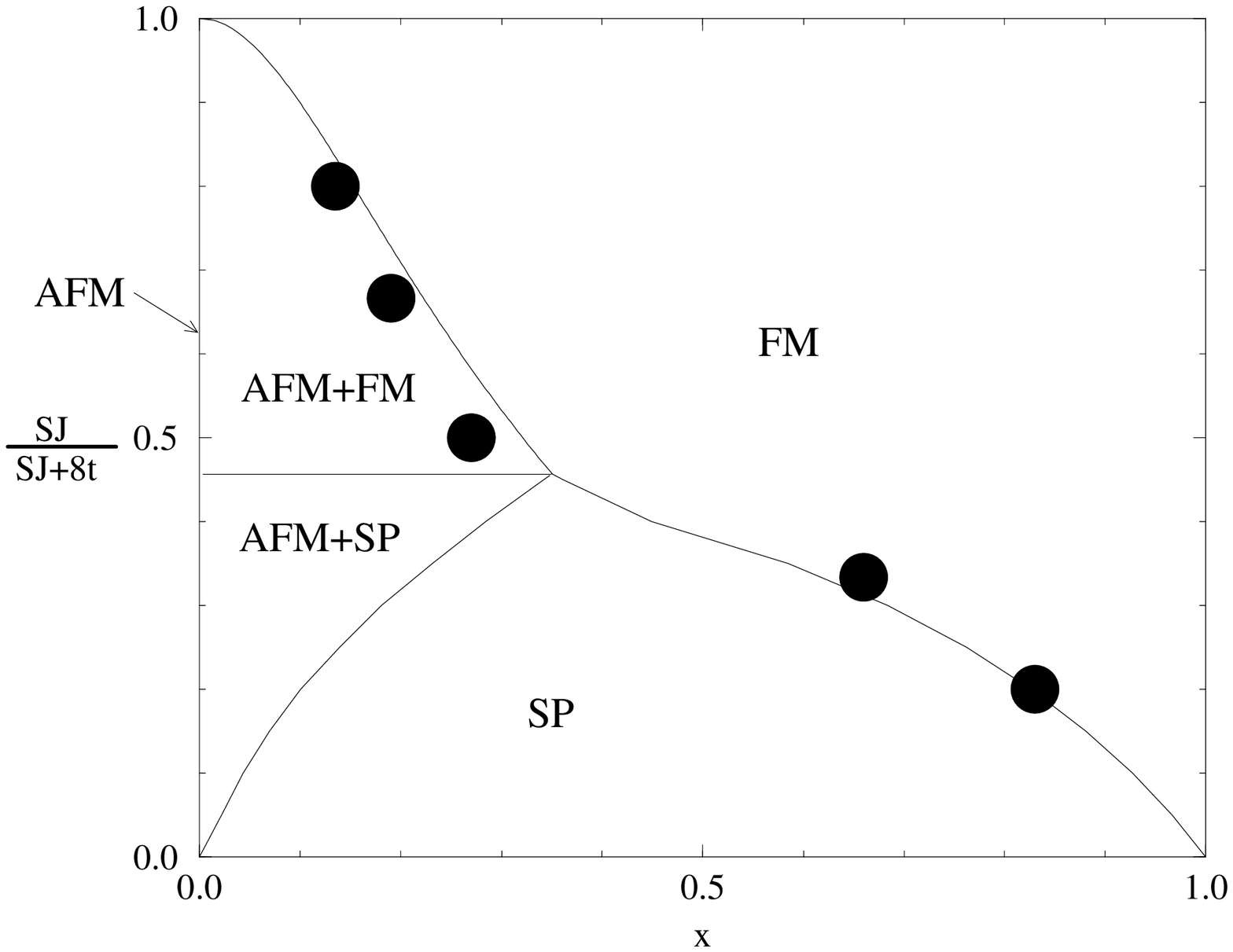} } \vspace{0.3cm}
\caption{\label{fig3} The zero temperature phase diagram of double
exchange model in square lattice in the large-S limit.  The black
spots are marks of of the phase boundaries taken from Ref.[6]. }

\end{figure}

The complete 3D and 2D phase diagrams are shown in Fig.~\ref{fig2} and
Fig.~\ref{fig3}.  Overall there are five
phases: the antiferromagnetic phase, the ferromagnetic phase, the spiral
phase, phase separation between the antiferromagnetic phase and the
ferromgnetic phase, and phase separation between the antiferromagnetic
phase and the spiral phase.  The phase transitions are likely to be second order
because both the energy derivative and the sizes of various phase regions
are continuous across the phase transition lines.  The 3D phase diagram
has a tricritical point at ${JS \over 4dt} \approx {2 \over 3}$, $x \approx 0.38$; the
tricritical point of 2D phase diagram is located at ${JS \over 4dt} \approx 0.84$,
$x \approx 0.35$.  However, the 3D and 2D phase diagrams has one major
difference.  In 3D, in the limit $J \rightarrow 0$,
the critical density between the ferromagnetic phase and the spiral phase
is 0.55; in 2D, it is unity.

In Fig.~\ref{fig3}, numerical data from Ref.~\cite{NUM} are also
plotted. The agreement is reasonably good considering that the
spin fluctuations could quantitatively modify our results and that
the finite size effect could affect the numerical results.
Although it is beyond the scope of our paper, we expect that the
spin fluctuations of the spiral states are described by
quadratically dispersed spin waves, similar to the ferromagnetic
case.  The dispersion should becomes more linear as doping
decreases.  It is very unlikely that the spin waves will make any
qualitative changes to the phase diagrams in 3D and 2D, although
more accurate numerical phase diagrams are also needed to make
comparison.

In one dimension however, quantum
fluctuations in principle destroy any broken symmetry states.  But
as found in Ref.~\cite{NUM}, phase diagrams of 1D systems are very
similar to 2D phase diagrams.  It is important to go beyond
the classical spin approximation to study 1D systems and the effects
of spin fluctuations.

In conclusion, we find that in the large-S limit of the DE model the
spiral phase is an intermediate phase between the ferromagnetic phase and
the antiferromagnetic phase when the Hund coupling is relatively small.
Near half filling, the system subjects to phase separation.  The 2D phase
diagram that we have found is consistent with the numerical
results~\cite{NUM}.  Future studies on more sophisticated DE model
is needed to relate to CMR systems.

The author would like to thank Sanjoy Sarker, Steven Kivelson and
especially Tin-Lun Ho for helpful discussions.  The author was
supported by the National Science Foundation under Grant No. DMR
0201948 during the stay in University of Washington, and is
currently supported by NSFC under Grant No. 10174003 and by SRF
for ROCS, SEM.

\begin{thebibliography}{99}
\bibitem{Intro}For introduction, see S. Jin, T. H. Tiefel, M. McCormack,
R. A. Fastnacht, and L. H. Chen, Science {\bf 264}, 413 (1994) and
references therein.
\bibitem{PD} P. Schiffer, A. P. Ramirez, W. Bao, and S-W. Cheong,
Phys. Rev. Lett. {\bf 75}, 3336 (1995); C.H. Chen and S-W. Cheong,
Phys. Rev. Lett. {\bf 76}, 4042 (1996).
\bibitem{DEM}C. Zener, Phys. Rev. {\bf 82}, 403 (1951); P. W. Anderson
and H. Hasegawa, {ibid}., {\bf 100}, 675 (1955).
\bibitem{DEFM}K. Kubo and N. Ohata, J. Phys. Soc. Jpn. {\bf 33},
21(1972); N. Furukawa, J. Phys. Soc. Jpn. {\bf 63}, 3214 (1994).
\bibitem{CAF}P. G. de Gennes, Phys. Rev. {\bf 118}, 141 (1960).
\bibitem{NUM}S. Yunoki, J. Hu, A. L. Malvezzi, A. Moreo, N. Furukawa, and
E. Dagotto, Phys. Rev. Lett. {\bf 80}, 845 (1998); E. Dagotto, S.
Yunoki, A. L. Malvezzi, A. Moreo, J. Hu, S. Capponi, D. Poilblanc,
and N. Furukawa, Phys. Rev. B {\bf 58}, 6414 (1998).
\bibitem{MNUM} A. Moreo, S. Yunoki, and E. Dagotto, Science {\bf 283}, 2034 (1999);
S. Yunoki and A. Moreo. Phys. Rev. B {\bf 58}, 6403 (1998); S.
Yunoki, A. Moreo, and E. Dagotto, Phys. Rev. Lett {\bf 81}, 5612
(1998); H. Yi and J. Yu, Phys. Rev. B {\bf  58}, 11123 (1998); J.
L. Alonso, J. A. Capit\'an, L. A. Fern\'andez, F. Guinea, and V.
Mart\'in-Mayor, Phys. Rev. B {\bf 64}, 054408 (2001).
\bibitem{FTPS}D. P. Arovas, G. G\'omez-Santos, F. Guinea, Phys. Rev. B {\bf 59},
13569 (1999).
\bibitem{TJM}S-Q. Shen and Z. D. Wang, Phys. Rev. B {\bf 58}, R8877 (1998).
\bibitem{ES}E. L. Nagaev, Phys. Rev. B {\bf 58}, 2415 (1998).
\bibitem{TP} N-H. Tong and F-C. Pu, Phys. Rev. B {\bf 62}, 9425 (2000).
\bibitem{CMD} A. Chattopadhyay, A. J. Millis, S. Das Sarma, Phys. Rev. B
{\bf 64}, 012416 (2001).
\bibitem{PSTJ} V. J. Emery, S. A. Kivelson, and H. Q. Lin, Phys. Rev. Lett. {\bf 64},
475 (1990).
\bibitem{SPR} B.I. Shraiman and E.D. Siggia, Phys. Rev. Lett. {\bf 62}, 1564 (1989);
C. Jayaprakash, H. R. Krishnamurthy, S. Sarker, Phys. Rev. B {\bf
40}, 2610 (1989).
\bibitem{DESPR} J. Inoue and S. Maekawa, Phys. Rev. Lett. {\bf 74}, 3407 (1998).
\bibitem{SB}S. Sarker, J. Phys. Cond. Mat. {\bf 8}, L515 (1996);
D.P. Arovas and F. Guinea, Phys. Rev. {\bf B 58}, 9150 (1998).
\end{thebibliography}
\end{document}